\documentclass[12pt]{article}
\usepackage{amsthm,amsmath,amsfonts,amssymb}
\usepackage{amsmath,amsfonts,amsthm}
\newcommand{\R}{{\mathord{\mathbb R}}}
\newcommand{\Z}{{\mathord{\mathbb Z}}}
\newcommand{\N}{{\mathord{\mathbb N}}}

\def\a   {\alpha}
\def\b   {\beta}
\def\Lam   {\Lambda}

\newcommand{\HH}{\mathcal{H}}
\newcommand{\DD}{\mathcal{D}}

\def\bN {\boldsymbol{N}}
\def\bmu {\boldsymbol{\mu}}
\def\bmut {\boldsymbol{\tilde{\mu}}}
\def\bepsilon {\boldsymbol{\epsilon}}
\def\brho {\boldsymbol{\rho}}

\def\tri {\triangle}
\def\triap {\triangle_{\alpha}^+}

\newcommand{\ben}{\begin{displaymath}}
\newcommand{\een}{\end{displaymath}}

\def\inf{{\rm inf}\,}

\def\Tr{\operatorname{Tr}}

\def\supp{\operatorname{supp}}

\newcommand{\derb}[2]{\frac{\partial #2 }{\partial #1}}

\newtheorem{lemma}{Lemma}
\newtheorem{theorem}[lemma]{Theorem}
\newtheorem{remark}{Remark}

\newtheorem{corollary}[lemma]{Corollary} 
\newtheorem{definition}{Definition}
\newtheorem{example}{Example}

\title{The Independence on Boundary Conditions for the Thermodynamic 
Limit of
Charged Systems
\thanks{Work 
partially supported by EU
grant HPRN-CT-2002-00277.}}
\author{\hspace{-.2 cm} David Hasler${}^{(a)}$, Jan Philip
Solovej${}^{(b)}$ \\
\normalsize\it \hspace{-.5 cm} \\
\hspace{-.5 cm}\normalsize\it${}^{(a)}$ Department of
Mathematics, University of British Columbia \\ 
\normalsize\it
Vanocuver B.C., Canada \\
\hspace{-.5 cm}\normalsize\it${}^{(b)}$ Department of
Mathematics, University of Copenhagen \\
\normalsize\it
2100 Copenhagen, Denmark }
\date{}

\begin{document}
\maketitle

\begin{abstract}
We study systems containing electrons and nuclei.
Based on the fact that the Thermodynamic limit exists for 
systems with Dirichlet boundary conditions, we prove that the same
limit is obtained if one imposes other boundary 
conditions such as Neumann, periodic, or elastic boundary 
conditions. The result is proven for all limiting sequences
of domains which are obtained by scaling a bounded open
set, with smooth boundary, except for isolated edges and corners. 
\end{abstract}

\section{Introduction}\label{sec:int}

We consider systems composed of electrons and nuclei, i.e.,
point particles which interact via Coulomb interaction with
the negatively charged particles being fermions.
Due to their important role in describing nature, 
such systems have been intensively investigated.
In particular, the thermodynamic limit, i.e.,  the limit in which the system
becomes large, has been studied extensively in
\cite{lieleb:the}. In that work, it was shown that
the thermodynamic limit exists for thermodynamic
quantities,  such as the pressure and the free energy density,
provided that they are defined using Dirichlet boundary conditions. 
Furthermore, it was shown that these quantities possess
the properties which are  expected from phenomenological
thermodynamics.

In order to define the canonical  and the grand
canonical partition function, one has to confine the particles
of the system
to lie in a bounded set $\Lambda \subset \R^3$, 
which we choose to 
be open. For the confined system to be well defined its 
Hamiltonian should be  self adjoint. This requires
that one imposes suitable boundary conditions on the 
boundary of $\Lambda$.
For each  particular choice of boundary conditions
one obtains a canonical and a grand canonical partition
function. In order to study the thermodynamic limit one considers
a  sequence
$\{ \Lambda_l \}$ of bounded
open domains such that the volume of $\Lambda_l$ tends
to infinity as $l \to \infty$. 
For systems with Dirichlet boundary conditions
it was shown in \cite{lieleb:the} that the 
canonical and grand canonical partition function exist
for a large class of limiting sequences $\{ \Lambda_l \}$. 
Moreover, the limit is independent of the particular sequence.

In this work we prove that, indeed, the same limit
is obtained for systems with Neumann, periodic, or 
reflecting boundary conditions.
We prove our result for limiting sequences which are 
obtained by scaling a bounded open set, which has 
a smooth boundary, except for isolated edges and corners.
This class of limiting sequences
is smaller than the class 
for which the thermodynamic limit 
for systems with Dirichlet boundary conditions has been shown to 
exist. 
We want to point out that this is only partially technical.
For instance, there exist sequences of domains for which the 
 thermodynamic limit of the 
ground state
energy for Dirichlet boundary conditions
exists, 
whereas for Neumann boundary conditions the ground state
energy  diverges to $- \infty$. 
Although such sequences are somewhat pathological, this demonstrates
that the independence of boundary conditions for  systems
composed of electrons and nuclei
cannot be considered as trivial. We will also comment on
possible more general classes of limiting sequences for which 
our proof is applicable. 
For notational simplicity, we only state and prove our results
for systems composed of a single species of negatively
charged fermions and a single species of positively charged
particles being bosons. The results as well as their proofs
generalize to multicomponent systems in a straight forward
way. 
We state the main result and present its proof
for both: zero temperature and nonnegative temperature.
Despite that the latter implies the former,
we present  that way  an independent and  
technically easier proof 
for the temperature
zero case.

To prove the independence of the boundary conditions we
use a sliding technique, which was introduced in 
\cite{conlieyau:the}, and refined in \cite{grasch:ont}.
Thereby, one decomposes the space into simplices.
By sliding and rotating the simplices one obtains 
a lower bound for the 
Hamiltonian of a large system in terms of 
Hamiltonians defined on the smaller simplices.
Simplices which lie in the interior of the large system 
have Dirichlet boundary conditions. Whereas simplices on the boundary,
i.e.,
simplices which intersect with the boundary of the large system,
are subject to  mixed boundary conditions.
Using that the many  body Coulomb potential can be estimated below by a sum of
one body potentials \cite{lieyau:the}, 
we then show that the  thermodynamic quantities in the boundary simplices 
are bounded.
In the thermodynamic limit the sum of all the boundary contributions 
is proportional to the surface. This is negligible
compared to  the bulk contribution, which is proportional to the
volume.  

We want to point out that independence
of boundary conditions has been studied 
for systems with hard core interactions (see \cite{rob:sta}, 
\cite{rob:the}, and references given therein). 

The paper is organized as follows. In Section \ref{sec:modres}
we introduce the model and state the results. In section \ref{sec:the}
we present the proofs.

\section{Model and Statement of Results}\label{sec:modres}

We shall first recall the definition of Dirichlet and 
Neumann boundary conditions \cite{reesim:ana}. Let $\Lambda$ be 
a bounded open set in $\R^3$. The Dirichlet Laplacian for $\Lambda$,
$- \Delta^{\rm D}_{\Lambda}$, is the unique self-adjoint operator
on $L^2(\Lambda)$ whose quadratic form is the closure of the 
form
\ben
\phi \mapsto
\int_{\Lambda} | \nabla \phi|^2 \, dx
\een
with domain $C_0^{\infty}(\Lambda)$. The Neumann Laplacian for $\Lambda$,
$-\Delta^{\rm N}_{\Lambda}$, is the unique self-adjoint operator on 
$L^2(\Lambda)$ whose quadratic form is 
\ben
\phi \mapsto
\int_{\Lambda} | \nabla \phi |^2 \, dx 
\een
with domain $H^1(\Lambda) = \{ \ f \in L^2(\Lambda) \ | \
\nabla f \in L^2(\Lambda) \ (\mathrm{in \ sense \ of \ distributions}) \  \}$.

The model consists of electrons ($\hbar^2/2 = 1$, $m=1$,
$|e|=1$) and nuclei with mass $M$ and charge $z$. We assume $z$ to be rational.
The electrons are
fermions, while the statistics of the nuclei is irrelevant. Let $\Lam
\in \R^3$ be an open set. The Hilbert space
$\HH_{\bN,\Lam}$, with $\bN=(n,k) \in \N^2$, for $n$
electrons and $k$ nuclei is the subspace of $L^2(\Lam \times
\Z_2)^{\otimes n} \otimes L^2(\Lam)^{\otimes k}$ carrying the
permutation symmetry appropriate to the given statistics. The
Hamiltonian, acting on  $\HH_{\bN,\Lam}$, is
\begin{eqnarray*}
H^B_{\bN,\Lam} & = & - \sum_{j=1}^n \Delta^B_{\Lam, x_j} - \frac{1}{M}
\sum_{j=1}^{k} \Delta^{B}_{\Lam, R_j} - z \sum_{i=1}^n
\sum_{j=1}^{k} \frac{1}{|x_i - R_j |}  \\
&   & + 
\sum_{1 \leq i < j \leq n} \frac{1}{|x_i - x_j |}  +
z^2 \sum_{1 \leq i < j  \leq k} \frac{1}{|R_i - R_j |} \; ,
\end{eqnarray*}
where the electron coordinates are $x_i$, the nuclear coordinates
are $R_i$ and by $B$ we denote the type of boundary conditions, e.g. 
${\rm N},{\rm D}$, and ${\rm M}$ stands for Neumann, 
Dirichlet, and mixed boundary
conditions. Variable 
particle numbers 
are accounted 
for by means of the direct sum
\begin{eqnarray}
\HH_{\Lam} 
 =    \bigoplus_{\bN} \HH_{\bN, \Lam}
  \label{sec:int:for1}     \ \ , \qquad 
H^B_{\Lam}  =  \bigoplus_{\bN} H^B_{\bN, \Lam} \; . 
\end{eqnarray}
The grand canonical partition function and the
(finite volume) pressure are defined by
\begin{eqnarray*}
\Xi^B (\b, \bmu, \Lam) & = & \Tr_{\HH_{\Lam}} e^{- \b ( H_{\Lam}^B - \bmu
    \cdot \bN) } 
= \sum_{\bN}  \Tr_{\HH_{\bN, \Lam}} e^{- \b ( H_{\bN, \Lam}^B - \bmu
  \cdot \bN)}   \\
p^B (\b, \bmu , \Lam ) & = & ( \b | \Lam | )^{-1} \log \Xi^B(\b, \bmu,
\Lam) \; ,
\end{eqnarray*}
where $\bmu = (\mu_n , \mu_k )
\in \R^2$ stands for the chemical potentials of the electrons and
the nuclei, $\b > 0$ is the inverse temperature, and $\bN$ denotes the
particle number operator, for which we 
use the same symbol as for its eigenvalues. 
Here and below the volume of a subset $\Omega$ in $\R^3$ is denoted 
by $|\Omega|$.  
The canonical partition function of the system at reciprocal
temperature $\b$ and the free energy per unit volume are defined by 
\begin{eqnarray*}
Z^B(\b,\bN,\Lam)&=& \Tr_{\HH_{\bN,\Lam}} e^{- \b H^B_{\bN,\Lam}} \\
f^B(\b , \bN , \Lam) &=& - ( \b |\Lam |)^{-1} \log Z^B(\b , \bN , \Lam ) \; .
\end{eqnarray*}
Furthermore, we consider the following zero temperature expressions, 
which we will denote as 
\ben
G^B(\bmu, \Lam) = \inf \sigma_{\HH_{\Lambda}} (H_{\Lam}^B - \bmu \cdot \bN ) \  , \quad
g^B(\bmu, \Lam) = \frac{1}{|\Lambda|} G^B(\bmu, \Lambda) \; ,
\een
\ben
E^B( \bN , \Lam ) = \inf \sigma_{\HH_{\Lambda}} (H^B_{\bN, \Lam})
\ , \quad  
e^B(\bN , \Lam ) = \frac{1}{|\Lam|} E^B(\bN,\Lam ) \; . 
\een

\begin{definition} 
A sequence $\{\Lambda_l\}$ of bounded open sets in  
$\R^3$ is called a regular sequence of
domains if:
\begin{itemize}
\item[(i)] For $l \to \infty$, $|\Lambda_l | \to \infty$.
\item[(ii)] For each fixed $h \geq 0$  
as $l \to \infty$ (with $\Lambda_l^c = \R^3 \setminus \Lambda_l$)
\\
\mbox{$
| \{ \ x \in \Lambda_l \ | \ d(x,\Lambda^c_l) < h \ \}| 
/|\Lambda_l|
\to 0  $}   \ \ and  \quad $
 | \{ \ x \in 
\Lambda_l^c \ | \ d(x,\Lambda_l) \leq  h \ \}| /|\Lambda_l|
\to 0   \; .
$
\item[(iii)] There exists a $\delta > 0$ such that for all $l$,
$| \Lambda_l | / | B_l | \geq \delta $,
where $B_l$ is the ball of smallest radius containing $\Lambda_l$.
\end{itemize}
\end{definition}

%
%


\vspace{0.3cm}

\noindent
It was shown in \cite{lieleb:the} that for regular sequences $\{
\Lam_l \}$
the thermodynamic limits
\ben
p^{\rm D}(\b, \bmu ) = \lim_{l \to \infty} p^{\rm D}( \b , \bmu , \Lam_l ) 
 \ , \qquad
\lim_{l \to \infty} g^{\rm D}(\bmu, \Lambda_l  ) = g^{\rm D}(\bmu)  
\een
for Dirichlet boundary conditions exist and are independent of the particular
sequence.
To study the thermodynamic limit for the canonical ensemble,
we consider systems with no net charge, i.e.,
\ben
\bN  =   ( n, k) \ \ \mathrm{with} \quad k =  z n  \; .
\een
We introduce the set
\ben
P_{S} = \{ \ (\rho_e, \rho_k,) \in \R_+^2 \ | \ \rho_k  = z \rho_e \ \}  
\; , 
\een
corresponding to neutral charge configurations.
In \cite{lieleb:the}, it was also shown
that for a regular sequence of domains
$\{ \Lam_l \}$ and neutral $\{ \bN_l \}$, i.e., $\bN_l  \in \N^2 \cap P_S$,
with 
\ben
\lim_{l \to \infty} \frac{\bN_l}{|\Lam_l|} = \brho \equiv (\rho_e ,
\rho_k)  \in P_S \; ,
\een
the limits for Dirichlet conditions
\ben
\lim_{l \to \infty} f^{\rm D}(\b , \bN_l , \Lam_l ) = f^{\rm D} ( \b ,\brho) \ , 
\qquad
\lim_{l \to \infty} e^{\rm D}(\bN_l , \Lam_l ) = e^{\rm D} (\brho)  \; 
\een
exist independent of the particular sequence
and are convex functions of $\brho \in P_S$. 
The value of $\brho$ gives the density of the particles.
Furthermore, it was shown that the
canonical and the grand canonical ensembles are equivalent, i.e.,
that
\begin{equation}
p^{\rm D}(\b ,\bmu ) = \sup_{\brho \in P_S} ( \brho \cdot \bmu - f^{\rm D}( \b ,
\brho) ) \; , \qquad 
g^{\rm D}(\bmu) = \inf_{\brho \in P_S} ( e^{\rm D}(\brho) - \bmu \cdot \brho )
\; .
\label{sec:int:eoe}
\end{equation}

\vspace{0.3cm}

\noindent
We note that (\ref{sec:int:eoe})  implies that $g^{\rm D}(\bmu)$
is  concave and that $p^{\rm D}(\b ,\bmu )$ is convex, and hence
they are continuous functions of $\bmu$. We now state our main result.

\begin{theorem}  \label{sec:int:thm1}
Let $\Lambda \subset \R^3$ be a bounded open set with smooth boundary,
except for isolated edges and corners. Consider
the sequence $\Lambda_L = L   \Lambda$ for $L > 0$.
Then
\begin{itemize}
\item[(a)] 
\quad $
g^{\rm N}(\bmu ) = \lim_{L \to \infty } | \Lam_L |^{-1} G^{\rm N}(\bmu, \Lam_L ) =
g^{\rm D}(\bmu)$.
\item[(b)] 
\quad 
$p^{\rm N} ( \b , \bmu ) = \lim_{L \to \infty} p^{\rm N}(\b , \bmu, \Lam_L )
 = p^{\rm D}(\b , \bmu )$.
\end{itemize} 
\end{theorem}

\vspace{0.5cm}
\noindent
We want to point out that   
sequences satisfying the assumption of  Theorem \ref{sec:int:thm1} 
are regular sequences of 
domains. Theorem  \ref{sec:int:thm1} has the following consequence.

\begin{corollary}  \label{sec:int:cor1}
Let $\{ \Lam_L \}$
be a sequence of domains as in Theorem  \ref{sec:int:thm1}.
Let $\{ \bN_L \}$ be a
sequence with  neutral charge configuration, i.e., 
$\bN_L \in \N^2 \cap P_S$, such that 
\ben
\lim_{L \to \infty} \frac{\bN_L}{| \Lam_L| } = \brho \in P_S \; .
\een
Then
\begin{itemize}
\item[(a)] \quad 
$
e^{\rm N}(\brho) := \lim_{L \to \infty} e^{\rm N}(\bN_L , \Lam_L ) = 
e^{\rm D}(\brho) \; ;
$
\item[(b)]  \quad 
$
f^{\rm N}(\b ,\brho ) := \lim_{L \to \infty} f^{\rm N}(\b, \bN_L , \Lam_L ) = f^{\rm D}(
\b , \brho ) \; .
$
\end{itemize}
\end{corollary}



\begin{remark}{\rm
Note that we only consider systems which consist of electrons
and one type of spinless nuclei. 
The results and their proofs generalize in a straight 
forward way to multicomponent systems, with all 
negatively (or positively) charged particles being fermions.}
\end{remark}

\begin{remark}{\rm
The essential technical requirement 
in the proof of Theorem \ref{sec:int:thm1} on the 
sequence of domains $\Lambda_L$, apart from being regular, is that
the thermodynamic quantities of the boundary simplices are bounded,
cp. Lemma \ref{sec:pro:lem2}.
This in turn
holds for all sequences satisfying the assertion of 
Lemma \ref{sec:pro:geom}, which is stated in the next section.
We want to point out
that there do exist sequences of domains for which
the thermodynamic limit does not exist
for systems with Neumann boundary
conditions, and yet for Dirichlet conditions the thermodynamic
limit exists, cp. the example below.}
\end{remark}


\begin{example}{\rm
We consider a system where the charge $z$ of the nuclei is one. 
Let $\{ \Lambda_l \}$ be the union of a large ball $B_l$ of radius $l$ and a 
shrinking
ball $B_{l^{-4}}$ of radius $l^{-4}$ separated from the large 
ball by a constant distance. 
Let $\{ \bN_l \}$  be a sequence with $\bN_l \in \N^2 \cap P_S$
 $\lim_{l \to \infty} | \Lambda_l |^{-1} \bN_l  = \brho$.
The sequence $\{ \Lambda_l \}$ is a regular sequence of domains.
We place one electron and a single nucleus in the small ball
and put both in the Neumann ground state. In that
situation the small ball has neutral
charge distribution and hence there is no Coulomb interaction with
the large ball. This provides us with the following upper bound
\ben
e^{\rm N}(\Lambda_l, \bN_l ) \leq |\Lambda_l|^{-1}
E^{\rm D}(B_l, \bN_l -(1,1))     
 + |\Lambda_l|^{-1} | B_{l^{-4}}|^{-2} \int_{(B_{l^{-4}})^2}
\frac{-1}{|x-y|} \, dx dy \; .   
\een
The first term on the right hand side converges to $e^{\rm D}(\brho)$
while the second term diverges to $-\infty$.
The same conclusion
is easily seen to hold 
if we connect the small ball to the large ball by a thin
tube provided that its thickness shrinks fast enough.}
\end{example}

\vspace{0.5cm}

Finally  we want to  consider more general boundary conditions.
Consider for instance, a Laplacian
$- \Delta^{\rm A}_{\Lam}$ with boundary conditions such that 
\begin{equation} \label{eq:perb}
- \Delta^{\rm N}_{\Lam} \leq - \Delta^{\rm A}_{\Lam} \leq - \Delta^{\rm D}_{\Lam} \;. 
\end{equation}
(Here and below operator inequalities are understood in 
the sense of forms
\cite{reesim:ana}.) Then Theorem 
\ref{sec:int:thm1} and Corollary
\ref{sec:int:cor1}, respectively,  imply that 
the same limits are obtained for systems 
with boundary conditions satisfying (\ref{eq:perb}). 
We note that periodic boundary conditions
are of this type.

Elastic boundary conditions, with elasticity $\sigma$, are 
defined as follows. Let $t^{\sigma}_{\Lambda}$ denote the quadratic
form
which is the closure of the form
\ben
\phi \mapsto \int_{\Lambda} | \nabla \phi |^2 \, dx 
+ \sigma \int_{\partial \Lambda} |\phi |^2 \, dS 
\een
with domain
$H^1(\Lambda) \cap C(\overline{\Lambda})$
and where  $dS$ is the surface measure of $\partial \Lambda$,
the boundary of $\Lambda$.
Let  $-\Delta^{\sigma}_{\Lambda}$ denote
the unique self adjoint operator with quadratic form $t^{\sigma}_{\Lambda}$. 
Functions in the domain of $- \Delta^{\sigma}_{\Lambda}$
satisfy
\ben
\left. 
\frac{\partial \phi}{\partial {n}} = \sigma \phi 
\right\vert_{\partial \Lambda} 
\een
at the boundary of $\Lambda$, where $\partial \phi / 
\partial {n}$ denotes the normal derivative. Note  that  boundary conditions
with 
elasticity zero are  Neumann boundary conditions.
For positive elasticity, $\sigma > 0$, we have the operator
inequality 
$
- \Delta^{\rm N}_{\Lambda} \leq  - \Delta^{\sigma}_{\Lambda}  \leq 
- \Delta^{\rm D}_{\Lambda}$.
This implies the statement of the following theorem in the case 
where the elasticity
is positive. That it, indeed, holds for negative elasticity
will be shown in Section \ref{sec:pro4}.

\begin{theorem} \label{sec:int:thm2}
Let $\Lambda_L$ be as in Theorem \ref{sec:int:thm1}. Then, for elastic boundary 
conditions with real elasticity $\sigma$, 
$\lim_{L \to \infty} p^{\sigma}(\beta , \bmu , \Lambda_L) = p^{\rm D}(\beta, \bmu)$.
Let $\{ \bN_L \}$ be a sequence as in Corollary \ref{sec:int:cor1}. Then
$\lim_{L \to \infty} f^{\sigma} (\beta , \bN_L , \Lambda_L ) = 
f^{\rm D}(\beta, \brho)$.
\end{theorem}

\section{Proofs} \label{sec:the}

First we show that Corollary   \ref{sec:int:cor1} follows 
from Theorem \ref{sec:int:thm1}. In subsection \ref{subsec:32}
we will prove Theorem  \ref{sec:int:thm1}, which is our main
result. The prove is based on a lemma which estimates the contributions
from the boundary terms. The prove of that lemma is deferred
to subsection \ref{subsec:33}.
In subsection \ref{subsec:34} we prove Theorem \ref{sec:int:thm2}
concerning reflecting boundary conditions.

\subsection{Proof of Corollary \ref{sec:int:cor1}}

\noindent
(a). We know that for  $\brho \in P_S$ 
\begin{equation} \label{sec:int:edr}
e^{\rm D}(\brho) 
\geq
\liminf_{L \to \infty} e^{\rm N}(\bN_L , \Lam_L ) \; .
\end{equation} 
For a given $\brho \in P_S$ 
there exists, by the convexity of $e^{\rm D}$, a $\bmu$
such that 
\ben
e^{\rm D}(\brho') \geq {e^{\rm D}(\brho) + \bmu \cdot ( \brho' - \brho)} 
\ \  , \quad \mathrm{for \ all} \  \brho' \in P_S \; .
\een
Hence
\begin{equation} \label{sec:int:ied1}
\inf_{\brho' \in P_S } ( e^{\rm D}(\brho') - \bmu \cdot \brho' ) =
e^{\rm D}(\brho) - \bmu \cdot \brho \; .
\end{equation}
We have
\begin{eqnarray*}
\liminf_{L \to \infty} e^{\rm N}(\bN_L , \Lam_L ) & = &  \liminf_{L \to
    \infty} \frac{E^{\rm N}(\bN_L,\Lam_L)}{|\Lam_L|} \\
& = & \liminf_{L \to \infty} \frac{E^{\rm N}(\bN_L , \Lam_L ) - \bmu \cdot  
\bN_L
}{|\Lam_L|} + \bmu \cdot \brho \\
& \geq & \liminf_{L \to \infty} \left( \inf_{\bN} 
 \, \frac{E^{\rm N}(\bN, \Lam_L ) - \bmu \cdot \bN }{|\Lam_L|} \right)  + 
\bmu \cdot \brho \\
& = & g^{\rm D}(\bmu) + \bmu \cdot \brho \\
& = & \inf_{\brho' \in P_S} ( e^{\rm D}(\brho') - \bmu \cdot \brho' ) +
\bmu \cdot \brho \\
& = & e^{\rm D}(\brho) \; ,
\end{eqnarray*}
where we have used Theorem \ref{sec:int:thm1} (a) in the fourth,
eq. (\ref{sec:int:eoe}) 
in the fifth and  eq. (\ref{sec:int:ied1}) in the last
line. The above  inequality together with  (\ref{sec:int:edr}) 
proves (a).

\vspace{0.3cm}
\noindent
(b). The proof of (b) is analogous to (a). We know that for 
$\brho \in P_S$
\begin{equation} \label{sec:int:fdb1}
- f^{\rm D}(\b , \brho ) 
\leq \limsup_{L \to \infty} ( - f^{\rm N}(\b , \bN_L , \Lam_L ) ) \; ,
\end{equation}
where we used that the map $A \mapsto \mathrm{Tr} \, e^A$ is operator 
monotone. 
Since $P_S \ni \brho \mapsto - f^{\rm D}(\b , \brho)$ is concave, 
there exists for a
given $\brho \in P_S$ a $\bmu$ such that
\ben
- f^{\rm D}(\b , \brho' ) \leq - f^{\rm D}(\b , \brho ) - \bmu \cdot ( \brho' -
\brho )   \ \  , \quad \mathrm{for \  all} \ \brho' \in P_S \; ,
\een
and hence
\begin{equation}   \label{sec:int:snr}
\sup_{ \brho' \in P_S} \left( \brho' \cdot \bmu - f^{\rm D}( \b , \brho')
\right) = \brho \cdot \bmu - f^{\rm D}( \b , \brho )  \; .
\end{equation} 
We have
\begin{eqnarray*}
\lefteqn{\limsup_{L \to \infty} \left( - f^{\rm N}(\b , \bN_L , \Lam_L ) \right) = }
\\ & & =
\limsup_{L \to \infty} \left( - \brho \cdot \bmu + (\b | \Lam_L
  |)^{-1} \b \bmu \cdot \bN_L  - f^{\rm N}(\b , \bN_L , \Lam_L ) \right) \\
& & = - \brho \cdot \bmu + \limsup_{L \to \infty} \left( ( \b | \Lam_L |)^{-1} 
 \log Z^{\rm N}(\b ,\bN_L , \Lam_L ) e^{ \b \bmu \cdot \bN_L } \right) \\
& & \leq - \brho \cdot \bmu + \limsup_{L \to \infty} \left( 
( \b | \Lam_L |)^{-1} 
 \log \left( \sum_{\bN} Z^{\rm N}(\b ,\bN , \Lam_L ) e^{ \b \bmu \cdot \bN }
 \right) \right)  \\
& & = - \brho \cdot \bmu + p^{\rm D}(\b , \bmu) \\
& & = - f^{\rm D}(\b , \brho) \; ,
\end{eqnarray*}
where we have used Theorem \ref{sec:int:thm1} (b) in the fourth and
eqns. (\ref{sec:int:eoe},\ref{sec:int:snr}) in
the last line. The above
inequality together with (\ref{sec:int:fdb1}) proves (b).
\qed

\subsection{Proof of Theorem  \ref{sec:int:thm1}}
 \label{subsec:32}

\noindent
To prove Theorem \ref{sec:int:thm1},
we will make use of the localization method of
\cite{grasch:ont} (see also \cite{conlieyau:the}), 
where one breaks up $\R^3$ into simplices in the following
way. Cutting the unit cube $W = [0,1]^3$ with all planes passing
through the centre and an edge or a face diagonal of $W$, one obtains
congruent simplices $\triangle_n \subset W$, ($n=1,...,24$). The
simplices $\tri_{\a} = \tri_n + z$, with $\a = (z,n) \in \Z^3 \times
\{ 1, ..., 24 \} = : I$, yield a partition of $\R^3$ up to their
boundaries. We then choose a spherically symmetric $\varphi_0 \in
C_0^{\infty}(\R^3)$ with $\int \varphi_0^2 = 1$ and $\{ \varphi_0(x)
\neq 0 \} = \{ |x| < 1/2 \}$. Let $\chi_{\a}$ be the characteristic
function of $\tri_{\a}$. Setting $\varphi(x) = \eta^{-3/2}
\varphi_0(x/\eta)$ and $j_{\a} = (\chi_{\a} * \varphi^2 )^{1/2}$, we
obtain a partition of unity, i.e.,
\ben
\sum_{\a \in I} j_{\a}^2(x) = 1 \ \ , \; ( x \in \R^3 ) \; ,
\een
with $j_{\a} \in C^{\infty}(\R^3)$. There are congruent simplices
$\tri_{\a}^+$, which are scaled copies of $\triangle_{\a}$, such that
\begin{eqnarray*}
& \supp j_{\a} \subset \triap & \\
& | \triap | \leq | \tri_{\a} | ( 1 + O(\eta)) &
\end{eqnarray*}
as $\eta \downarrow 0$.

The following definitions depend on $\eta$ and $l > 0$ although the
notation will not reflect this for simplicity. 
For the moment, let $\Lam \subset \R^3$ be any bounded open set.
For $y \in W$ and $R \in SO(3)$ we set
\ben
\Lam^{y,R} = R^{-1} \Lam - l y \; .
\een
We define the set 
\ben
I(\Lam) = \{ \a \in I | l \triap  \cap \Lam^{y,R}  \neq
\emptyset \ \mathrm{for \ some} \ y \in W, R \in SO(3) \} \; .
\een
For $\a \in I(\Lam)$, let $\HH_{\a} := \HH_{l \triap}$ be the 
many particle space for the simplices $l \triap$ as given in
(\ref{sec:int:for1}). By $H^{\rm M}_{\a,y,R,\Lam}$ we denote the Hamiltonian
on $l \triap \cap \Lam^{y,R}$ with Neumann conditions on $(l
\triap) \cap \partial \Lam^{y,R}$ and Dirichlet conditions on the remaining
part of the boundary. The operator
$H^{\rm M}_{\a,y,R,\Lam}$  acts on 
\ben
\HH_{\a , y, R, \Lam} := \HH_{l \triap \cap \Lam^{y,R}}
\hookrightarrow \HH_{\a}
\een
and hence on $\HH_{\a}$ via the canonical embedding. Note, if $l\triap
\subset \Lam^{y,R}$, then $H^{\rm M}_{\a,y,R,\Lam} = 
H^{\rm D}_{l \triangle_{\a}^+}$ and $\HH_{\a,
  y, R, \Lam} = \HH_{\a}$. 
We define the Hilbert space and a Hamiltonian, acting on it, as the
direct integrals
\begin{eqnarray*}
\HH_{I(\Lam)} & = & \int^{\oplus}_{W \times
  SO(3)} dy d \mu (R)  \bigotimes_{\a \in I(\Lam)} \HH_{\a} \; , \\
H^{\rm N}_{I(\Lam)} & = & \int^{\oplus}_{W \times SO(3)} d y d \mu(R) 
\sum_{\a \in I(\Lam) } H^{\rm M}_{\a ,y, R, \Lam } \; ,
\end{eqnarray*}
where $d \mu $ denotes the Haar measure on $SO(3)$. We shall define a
map $J : \HH_{\Lam} \to \HH_{I(\Lam)}$ as follows. Let 
$j_{y,R,\a} : L^2(\Lam) \to L^2(l \triap)$ be given by
\ben
( j_{y,R,\a} \psi)(x) = j_{\a}(x/l) \psi( R(x + l y)) \; .
\een
Define
\begin{eqnarray*}
j_{y,R} : L^2(\Lam) & \to & \bigoplus_{\a \in I(\Lam)} L^2(l \triap
 )
\\
j_{y,R} & = & \bigoplus_{\a \in I(\Lam)} j_{y,R,\a} \; .
\end{eqnarray*}
This lifts to a  map between the many particle spaces
\ben
\Gamma(j_{y,R} ) : \HH_{\Lambda}  \to \bigotimes_{\a \in I(\Lam)}
\HH_{\a} \; ,
\een 
which acts as the $\bN$-fold tensor product of $j_{y,R}$ on $\bN$-particle
states. 
We may now define
\ben
J : \HH_{\Lam} \to \HH_{I(\Lam)} \; , \qquad J = \int_{W \times
  SO(3)}^{\oplus} dy d \mu(R) \Gamma(j_{y,R}) \; .
\een 
We note that the map $j^*_{y,R,\a} : L^2(l \triap  ) \to
L^2(\Lam)$ is 
given by
\ben
(j^*_{y,R,\a} \psi)(x) = j_{\a} (R^{-1}(x/l) - y) \psi(R^{-1}x - l y )
\; .
\een 
Hence $j^*_{y,R} j_{y,R} : L^2(\Lam) \to L^2(\Lam)$ acts as
multiplication
by $\sum_{\a \in I(\Lam)} j_{\a}^2(R^{-1}(x/l) - y )$. This function
of $x \in \Lam$ equals $1$. We conclude that $J^* J =1$, i.e., that
$J$ is an isometry. We state the following Lemma, c.f. Lemma 7  in
\cite{grasch:ont}.

\begin{lemma}  \label{sec:pro:lem1}
Let $\eta = l^{-1}$. Then
\ben
H_{\Lam}^{\rm N} \geq \kappa J^* H_{I(\Lam)}^{\rm N} J - l^{-1} \mathbf{const}
\cdot \bN
\een
for large $l$, where $0 < \kappa \leq 1$ and $\kappa = 1 + O(l^{-1})$
as $l \to \infty$.
\end{lemma}
For the proof of this Lemma we refer the reader to the proof of
Lemma 7 in \cite{grasch:ont}, where the  statement is for the Dirichlet
Laplacian. With little modification of the proof given there one can
proof Lemma \ref{sec:pro:lem1}.

From now on, let $\Lambda$ be a fixed open set as  in the assumption
of Theorem 
\ref{sec:int:thm1},
i.e., bounded with smooth boundary, except for isolated edges and   
corners. We will consider the sequence of scaled copies $\Lambda_L = 
L \Lambda$, with $L > 0$. 
Note that the claim of Lemma \ref{sec:pro:lem1} of course
holds for all $\Lambda_L$.
Let $\triangle$ denote a simplex which is similar to one (and thus all)
of the simplices
$\triangle_{\alpha}$, i.e., equal up to dilations, translations, 
and rotations. 
By $\triangle_c$ we denote  its intersection with $\Lambda_L$, i.e.,
$\triangle_c = \triangle \cap \Lam_L$. Let  $-\Delta^{\rm M}_{\triangle_c}$
$(H^{\rm M}_{\triangle_c})$ denote the Laplacian (Hamiltonian)
on $\triangle_c$ with  
Neumann conditions 
on $\triangle \cap \partial \Lam_L$ and Dirichlet conditions
on the rest of $\partial \triangle_c$. 
We note that $-\Delta^{\rm M}_{\triangle_c}$ is the unique self adjoint
operator on $L^2(\triangle_c)$ whose quadratic form is the closure
of the form $\phi \mapsto \int_{\triangle_c} | \nabla \phi |^2 \, dx$
with domain $\{ \, \phi \in H^1(\triangle_c) \cap C(\overline{\triangle_c}) \, |
\, \phi \ \mathrm{vanishes \ in \ a \ neighborhood \ of} \  \partial
\triangle
\cap  \Lambda_L \, \}$.

The following Lemma whose prove will be postponed to subsection
\ref{subsec:33} 
provides us with a bound for the contributions from the boundary simplices.

\begin{lemma} \label{sec:pro:lem2} Let $\Lam_L$ be a sequence of domains 
as in Theorem \ref{sec:int:thm1} and 
let $v > 0$. Then there exists a number $L_0$ 
and constants   $C_E( \bmu, v )$ and 
 $C_{\Xi}(\b , \bmu, v)$ 
such that
\begin{itemize}
\item[(a)] 
\quad $
G^{\rm M}(\bmu, \triangle_c ) 
\geq C_E( \bmu, v ) >
  - \infty \;   ;
$
\item[(b)]
\quad $
\Tr_{\HH_{\triangle_c}} e^{ - \b ( H^{\rm M}_{\triangle_c} -  
\bmu \cdot \bN )}
 \leq
C_{\Xi}(\b , \bmu, v) < \infty \; ,
$
\end{itemize}
 for all
$\triangle_c =   \triangle \cap \Lambda_L$, 
with  $L \geq L_0$ and 
$|\triangle| \leq v$.

\end{lemma}

\vspace{0.5cm}
\noindent
For the proof of Theorem \ref{sec:int:thm1} 
we will also use the following lemma.

\begin{lemma} \label{sec:pro:lem4}
Let $\{ \bmu_l \} $ be a sequence in $\R^2$ with $\lim_{l
\to \infty} \bmu_l = \bmu$ and let $\{ \Lam_l \}$ be a regular 
sequence of domains.  
Then
\begin{itemize}
\item[(a)] 
\quad $
\lim_{l \to \infty} |\Lam_l|^{-1} G^{\rm D}(\bmu_l, \Lam_l )= 
\lim_{l \to \infty} |\Lam_l|^{-1} G^{\rm D}(\bmu, \Lam_l )= 
g^{\rm D}(\bmu) \;$  ;
\item[(b)]
\quad  
$
\lim_{l \to \infty} p^{\rm D}(\b, \bmu_l, \Lam_l ) =
\lim_{l \to \infty} p^{\rm D}(\b, \bmu, \Lam_l )  =
 p^{\rm D}(\b, \bmu ) \; .
$
\end{itemize}
\end{lemma}

\begin{proof}
We use the notation $\bmu_l = ({\mu_n}_l , {\mu_k}_l)$ and
$\bepsilon=(\epsilon, \epsilon)$.
For $\epsilon > 0$, there exists an $l_0$ such that for all $l \geq l_0$
\begin{eqnarray*}
\mu_n - \epsilon  \leq &{\mu_n}_l& \leq \mu_n + \epsilon  \\
\mu_k - \epsilon \leq &{\mu_k}_l& \leq \mu_k + \epsilon \; .
\end{eqnarray*} 
For (a), we note that
\ben
|\Lam_l|^{-1} G^{\rm D}(\bmu + \bepsilon, \Lam_l ) \leq 
|\Lam_l|^{-1} G^{\rm D}(\bmu_l, \Lam_l ) \leq 
|\Lam_l|^{-1} G^{\rm D}(\bmu - \bepsilon, \Lam_l ) \; , \qquad \forall \ l \geq l_0
\; .
\een
Hence 
\ben
g^{\rm D}(\bmu + \bepsilon) \leq \liminf_{l \to \infty} |\Lam_l|^{-1}
G^{\rm D}(\bmu_l, \Lam_l ) 
\leq 
\limsup_{l \to \infty} |\Lam_l|^{-1} G^{\rm D}(\bmu_l, \Lam_l ) \leq 
g^{\rm D}(\bmu - \bepsilon) \; ,
\een
and, by the continuity of $\bmu \mapsto g^{\rm D}(\bmu)$, (a) follows.

For (b), we first note that, by equation (\ref{sec:int:eoe}),
$\bmu \mapsto p^{\rm D}(\b , \bmu)$ is convex
and hence continuous. In analogy to (a) we have, using
that $A \mapsto \mathrm{Tr} \, e^A$ is (operator) monotone,
\ben
p^{\rm D}(\b, \bmu - \bepsilon, \Lam_l ) \leq 
p^{\rm D}(\b, \bmu_l, \Lam_l ) \leq 
p^{\rm D}(\b, \bmu + \bepsilon, \Lam_l ) \; , \qquad \forall \ l \geq l_0
\; .
\een
Hence 
\ben
p^{\rm D}(\b, \bmu - \bepsilon) \leq \liminf_{l \to \infty} p^{\rm D}(\b, \bmu_l, \Lam_l )
\leq 
\limsup_{l \to \infty} p^{\rm D}(\b , \bmu_l, \Lam_l ) \leq 
p^{\rm D}(\b ,\bmu + \bepsilon) \; ,
\een
and (b) follows by the continuity of  $\bmu \mapsto p^{\rm D}(\b , \bmu)$.
\end{proof}

\vspace{0.3cm}
\noindent
{\it Proof of Theorem \ref{sec:int:thm1}.} 
(a) Since $H_{\Lam_L}^{\rm D} - \bmu \cdot \bN \geq H_{\Lam_L}^{\rm N} - \bmu
\cdot \bN$, the inequality 
\begin{equation} \label{eq:ineq1}
 g^{\rm D}(\bmu) \geq \limsup_{L \to \infty} g^{\rm N}(\bmu, \Lam_L)
\end{equation}
is obvious.
We shall show the inequality 
$\liminf_{L \to \infty} g^{\rm N}(\bmu, \Lam_L) \geq g^{\rm D}(\bmu)$. 
We introduce  
\ben
\bN_{I(\Lam)} = \int_{W \times SO(3)}^{\oplus} dy d\mu(R) \sum_{\a \in
  I(\Lam)} \bN_{\a} \; ,
\een
where $\bN_{\a}$ denotes the number operator of 
$\HH_{\a}$. Note that $J^* \bN_{I(\Lam)} J = \bN$. By Lemma
\ref{sec:pro:lem1},
\ben
H_{\Lam}^{\rm N} - \bmu \cdot \bN \geq \kappa J^* \left( H_{I(\Lam)}^{\rm N} - \bmut_l
\cdot \bN_{I(\Lam)} \right) J  \; ,
\een
where we have set $\bmut_l = (1/\kappa) (\bmu + l^{-1}\mathbf{const}
)$. We define 
\begin{eqnarray*}
I^{\rm int}_{y,R}(\Lambda) & = & \{ \a | l \triap  \subset
 \Lam^{y,R}  \} \\
I^{\rm b}_{y,R}(\Lambda) & = &  
\{ \a | l \triap  \cap  \partial \Lam^{y,R}  \neq \emptyset \} \; .
\end{eqnarray*}
Let $\Psi \in \HH_{\Lam_L}$ be  normalized to one and smooth. We
observe that 
\begin{eqnarray*}
\lefteqn{
\left( \Psi, ( H^{\rm N}_{\Lam_L} - \bmu \cdot \bN ) \Psi
\right) } \\ 
& \geq & \kappa
\int_{W \times SO(3)} dy d \mu (R) 
\left( \Gamma(j_{y,R}) \Psi, \sum_{\a \in I(\Lam_L)} 
(H_{\a,y,R,\Lam}^{\rm M} - \bmut_l
\cdot \bN_{\a} ) \Gamma(j_{y,R}) \Psi \right) \\
& \geq & \kappa \int_{W \times SO(3)} d y d \mu (R) 
\left( 
\sum_{  \a \in I^{\rm int}_{y,R}  (\Lambda_L)  } G^{\rm D}(\bmut_l, l \triap) +
\sum_{ \a \in I^{\rm b}_{y,R}   (\Lambda_L)   }
G^{\rm M}(\bmut_l, l \triap  \cap \Lam_L^{y,R})
\right) \; .
\end{eqnarray*}
Hence 
\begin{eqnarray*}
\lefteqn{ \liminf_{L \to \infty} |\Lam_L|^{-1} G^{\rm N}(\bmu,\Lam_L) 
 \geq } \\ 
& & \liminf_{L \to \infty} \kappa \int_{W \times SO(3)} d y d \mu (R)
\left(
\sum_{    \a \in I^{\rm int}_{y,R}  (\Lam_L)     } \frac{| l \triap |}{|
  \Lam_L |} \cdot | l \triap |^{-1} G^{\rm D}(\bmut_l, l \triap )  \right. \\
& & \hspace{7cm}  + \left.
\sum_{   \a \in I^{\rm b}_{y,R}   (\Lam_L)    } 
 \frac{1}{| \Lam_L |}  C_E(\bmut_l, | l\triap |) 
\right)    \\
& & \geq \kappa ( 1 + O(l^{-1})) \cdot | l \triangle^+ |^{-1} 
G^{\rm D}(\bmut_l, l \triangle^+ ) \; ,
\end{eqnarray*}
where we used that all simplices $\triap$ are congruent to 
a single one, which we denote by $\triangle^+$, and  
in the last inequality we 
used Lemma \ref{sec:pro:lem2} (a) and that both limits
\begin{eqnarray} \label{sec:pro:for1}
 \lim_{L \to \infty}  \sum_{ \a \in I^{\rm int}_{y,R}  (\Lam_L)
} \frac{| l \triap |}{| \Lam_L |} & = & 1 + O(l^{-1}) \; , \\
 \lim_{L \to \infty} 
\sum_{ \a \in I^{\rm b}_{y,R}   (\Lam_L)     } 
 \frac{1}{| \Lam_L |} & = & 0 \;   \label{sec:pro:for2}
\end{eqnarray}
are uniform in $R \in SO(3)$, $y \in W$. We omit a proof of this
simple facts.
By Lemma \ref{sec:pro:lem4} (a),  the subsequent limit $l \to \infty$ yields
\begin{equation} \label{eq:ineq2} 
\liminf_{L \to \infty} g^{\rm N}(\bmu, \Lam_L) \geq g^{\rm D}(\bmu) \;.
\end{equation}
The two inequalities (\ref{eq:ineq1}) and (\ref{eq:ineq2}) show the 
claim.

(b) The inequality 
\begin{equation} \label{eq:ineq3}
p^{\rm D}(\b , \bmu) \leq 
\limsup_{L \to \infty} p^{\rm N}(\b , \bmu, \Lambda_L )
\end{equation}
 follows from
$-\Delta^{\rm N} \leq - \Delta^{\rm D}$. The opposite inequality 
\ben
\limsup_{L \to \infty} p^{\rm N}(\b , \bmu , \Lambda_L )  \leq   p^{\rm D}(\b , \bmu) 
\een
is seen as
follows. 
We set $\bmu_l = \bmu + l^{-1}
\mathbf{const}$. 
Let $\{ \varphi_i \}_{i \in I}$ be an eigenbasis of $J^* ( \kappa 
H_{I(\Lam_L)} - \bmu_l N_{I(\Lam_L)} ) J$. 
Then, using 
Lemma \ref{sec:pro:lem1}, we have 
\begin{eqnarray*}
\Xi^{\rm N}(\b, \bmu, \Lam_L) & \leq & \Tr_{\HH_{\Lam_L}} e^{- \b J^* (\kappa
  H_{I(\Lam)} - \bmu_l \cdot \bN_{I(\Lam_L)} ) J } \\
& = & \sum_{i \in I} e^{- \b ( J \varphi_i , (  \kappa H_{I(\Lam_L)} - \bmu_l
    \cdot \bN_{I(\Lam_L)} ) J \varphi_i )} \\
& \leq & \sum_{i \in I} \left( J \varphi_i , e^{- \b  ( \kappa H_{I(\Lam_L)} - \bmu_l
    \cdot \bN_{I(\Lam_L)} )} J \varphi_i \right) \\
& \leq  & \Tr_{\HH_{I(\Lam_L)}} e^{- \b ( \kappa H_{I(\Lam_L)} - \bmu_l
    \cdot \bN_{I(\Lam_L)} ) }   \\
& \leq & \int_{W \times SO(3)} dy d \mu (R) 
\prod_{    \a \in I^{\rm int}_{y,R}  (\Lam_L)        }
\Tr_{\HH_{\a}} e^{- \b 
  ( \kappa H^{\rm D}_{\a} - \bmu_l     \cdot   \bN_{\a} )} \\
&  & \qquad \times
\prod_{   \a \in I^{\rm b}_{y,R}   (\Lam_L)        } 
 \Tr_{\HH_{\a,y,R,\Lam_L}} e^{- \b ( \kappa H^{\rm M}_{\a,y,R,\Lam_L} - \bmu_l
    \cdot \bN_{\a} )} \; ,
\end{eqnarray*}
where, in the third line, we used Jensen's inequality with the spectral
measure of  $(\kappa H_{I(\Lam_L)} - \bmu_l
 \cdot \bN_{I(\Lam_L)})$   for $J \varphi_i$. 
Since $0 < \kappa \leq 1$, we have $\kappa H^{\rm D}_{\a} \geq \kappa^2 T^{\rm D}_{\a} +
\kappa V \cong H^{\rm D}_{\kappa^{-1}l \triap}$, where $T^{\rm D}_{\a}$ denotes the
kinetic Energy, $V$ is the Coulomb potential, and the unitary equivalence
comes from scaling. Note that all  the simplices $\triap$ are 
congruent to a single
one $\triangle^+$.
By Lemma \ref{sec:pro:lem2} (b), we have 
\begin{eqnarray*}
\lefteqn{p^{\rm N}(\b, \bmu, \Lam_L )} \\
&  =  & (\b | \Lam_L |)^{-1} \log \Xi^{\rm N} (\b
  , \bmu, \Lam_L ) \\ 
 & \leq & ( \b |\Lam_L |)^{-1} 
\log 
 \left( \Xi^{D} (\b , \bmu_l , \kappa^{-1} l \triangle^{+})^{\sup_{y,R}(|
  I^{\rm int}_{y,R}  (\Lam_L)  |)}  \cdot C_{\Xi}(\kappa \b, \kappa^{-1}\bmu_l, 
| l \triangle^+ | )^{\sup_{y,R}(|I^{\rm b}_{y,R}  (\Lam_L)|)} \right) \\
& \leq & \sup_{y,R} ( | I^{\rm int}_{y,R}    (\Lam_L) |) | \Lam_L |^{-1}
|\kappa^{-1} l \triangle^+ |   
\cdot p^{\rm D}(  \b ,  \bmu_l , \kappa^{-1} l \triangle^+ )  \\
& &  \ + \sup_{y,R}( |I^{\rm b}_{y,R}  (\Lam_L)|)
 ( \b |\Lam_L |)^{-1}  \cdot \log C_{\Xi}(\kappa \b, \kappa^{-1} \bmu_l,  
|l \triangle^+ | )  \; ;
\end{eqnarray*}
note that $\Xi^{\rm D} \geq 1$ and $C_{\Xi} \geq 1$. 
Thus
\ben
\limsup_{L \to \infty} 
p^{\rm N}( \b, \bmu , \Lambda_L )
 \leq  \kappa^{-3} (1 + O(l^{-1})) p^{\rm D}(\b , \bmu_l , \kappa^{-1} l 
\triangle^+ ) \; ,
\een
where we have used equations (\ref{sec:pro:for1}, \ref{sec:pro:for2}).
Using Lemma \ref{sec:pro:lem4} (b), the subsequent limit $l \to
\infty$ gives 
\begin{equation}  \label{eq:ineq4}
\limsup_{L \to \infty} 
p^{\rm N}(\b, \bmu ) \leq p^{\rm D}(\b , \bmu ) \; .
\end{equation} 
The claim in (b) follows from  eqns. (\ref{eq:ineq3}) and  
(\ref{eq:ineq4}).
\qed

\vspace{0.3cm}

\vspace{0.3cm}

\subsection{Proof of Lemma \ref{sec:pro:lem2}}
\label{subsec:33}

To prove Lemma \ref{sec:pro:lem2}, we 
will first state a technical Lemma 
reflecting the geometry of $\Lambda$. We recall that $\triangle$
denotes a simplex which is similar to one of the 
$\triangle_{\a}$, i.e., equivalent up to dilations, translations
and rotations.

\begin{lemma}  \label{sec:pro:geom}
{\rm \bf }  
Let $\Lambda_L$ be a sequence of domains as in Theorem \ref{sec:int:thm1}
and let $v>0$.
Then there exist a constant $C_{\Lambda} > 0$
and a number  $L_0$ (depending on $v$) 
such that for all $L \geq L_0$ and
all simplices $\triangle$ with $|\triangle| \leq v$ which intersect 
$\partial \Lambda_L$, we can choose an open set $V$ containing
$\triangle_c = \triangle \cap \Lambda_L$ and smooth coordinates 
\ben
\varphi :  V \to B_0 \quad  x \mapsto \varphi(x) = ( y_1(x),y_2(x), y_3(x))  
\een
with the properties:
\begin{itemize}
\item[(i)] $B_0$ is a ball centered around the origin
and $\Lambda \cap  V$
corresponds to either the half space restriction  
$\{x\in V \ |\ y_3(x)<0 \}$, the quarter 
space restriction $\{x\in V \ |\ y_2(x),y_3(x)<0 \}$, or the 
octant restriction $\{x\in V \ |\ y_1(x), y_2(x),y_3(x)<0 \}$. 
\item[(ii)] The Jacobian $D \varphi$ has determinant one
and $D \varphi^{-1} (D \varphi^{-1})^T  \geq C_{\Lambda}$.
\end{itemize} 
\end{lemma}

We want to point out that in the case where $\Lambda$ is a 
box, this Lemma follows trivially by choosing $C_{\Lambda}=1$
and the coordinate
maps to be an appropriate composition of a translation
followed by a rotation. In that case, also the next lemma
is trivial. 

\vspace{0.3cm}

\noindent
{\it Proof.} 
By assumption, $\Lambda$ is a bounded  subset  of $\R^3$ with smooth
boundary, except for isolated edges or corners. This means that 
around any point $x_0 \in \partial \Lambda$ on the boundary of 
$\Lambda$ there is an open neighborhood $V_{x_0}$ on which we may choose
smooth coordinates $(y_1,y_2,y_3)$ such that $\Lambda \cap  V_{x_0}$
corresponds to either the half space restriction  
$\{x\in V_{x_0}\ |\ y_3(x)<0\ \}$, the quarter 
space restriction $\{x\in V_{x_0}\ |\ y_2(x),y_3(x)<0\ \}$, or the 
octant restriction $\{x\in V_{x_0}\ |\ y_1(x), y_2(x),y_3(x)<0\ \}$.  
By rescaling a coordinate we can achieve  that the 
coordinate map $\varphi$ has Jacobian determinant equal to 
one\footnote{This    
can be achieved as follows. Let $\phi : x \mapsto       
w(x) = (w_1(x),w_2(x),w_3(x))$ be a coordinate map 
with Jacobian  
determinant 
not necessarily equal to one. Then, 
we define new coordinates    
\ben   
y_3(w) = \int_0^{w_3} | \det D \phi^{-1}(w_1,w_2,s) | \, ds \ \ , 
\quad y_1 = w_1 \ , \ 
y_2 = w_2 \; . 
\een  
It follows that 
$dy = | \det D \phi^{-1} |  \, dw$
and thus the Jacobian determinant    
of the coordinate map $x \mapsto y(w(x))$  is 1.}. 
By possibly adjusting the coordinates and  
choosing the neighborhood $V_{x_0}$ smaller, we can achieve 
that the images  
of the coordinate neighborhoods $V_{x_0}$ under the coordinate maps are   
balls centered at the origin.   
By compactness there exist constants $C_{\Lambda} > 0$ and $r > 0$
such that for each point $x_0 \in \partial \Lambda$  
we can choose a coordinate map 
$\varphi : V_{x_0} \to B$ such that 
\ben
 D \varphi^{-1} ( D\varphi^{-1})^T  \geq C_{\Lambda} \; 
\een
and moreover $|x - x_0 | < r$ implies $x \in V_{x_0}$. 
%
Given such a collection of coordinate maps for $\Lambda$ we obtain,
by scaling, 
a collection of coordinate charts for $\Lambda_L$ with
properties {\it(i)} and {\it(ii)}.
Moreover, the constant $r$ becomes $L r$ under this scaling. Thus for 
large $L$, $\triangle_c = \triangle \cap \Lambda_L$ 
is contained in some coordinate chart.
\qed

\vspace{0.5cm}

Let  $-{\Delta}^{\rm M}_{\varphi (\triangle_c)}$ denote the Laplacian
on $\varphi ( \triangle_c)$ 
with mixed boundary conditions, i.e., $\varphi$ maps Dirichlet (Neumann)
boundaries of $\triangle_c$ to Dirichlet (Neumann) boundaries of 
$\varphi(\triangle_c)$. 

\begin{lemma} \label{sec:pro:ineq} Let $\varphi$ be a coordinate map as in 
Lemma  \ref{sec:pro:geom}. Then the map
$ U : L^2(V) \to L^2(B_0)$  $f \mapsto f \circ \varphi^{-1}$
is unitary. Moreover,  on the form domain of $ - \Delta^{\rm M}_{\triangle_c}$
\ben
- \Delta^{\rm M}_{\triangle_c}  
\geq U^* C_{\Lambda} ( - \Delta^{\rm M}_{\varphi(\triangle_c)}) U
\; .
\een
\end{lemma}

\vspace{0.3cm}
\noindent
{\it Proof.}
Since the Jacobian determinant of $\varphi$ is one, $U$ is unitary.
By abuse of notation we write $f(y)$ for
$(U f ) (y) = f \circ \varphi^{-1}(y)$. 
We set  $g^{ij} = ( D\varphi^{-1}  ( D \varphi^{-1})^{T} )_{ij}$.
For functions $f$
in the form domain of $- \Delta^{\rm M}_{\triangle_c}$ we write
$( f ,  - \Delta^{\rm M}_{\triangle_c} f ) $
in terms of the $y$ coordinates and estimate
\begin{eqnarray*}
( f ,  - \Delta^{\rm M}_{\triangle_c} f ) 
&&= \int_{\varphi(\triangle_c)} \sum_{i,j} g^{ij}  
\overline{\derb{y_i}{f}} \hspace{-1.5mm}(y) 
  \derb{y_j}{f}(y)  \, dy \\
&& \geq \int_{\varphi(\triangle_c)}  
\sum_{i,j } C_{\Lambda} \delta^{ij} \overline{\derb{y_i}{f}} \hspace{-1.5mm}(y)  
\derb{y_j}{f}(y) 
  d y  \\
&& = C_{\Lambda} ( Uf ,  -  {\Delta}^{\rm M}_{\varphi(\triangle_c)} Uf ) \; .
\end{eqnarray*}
\qed

\begin{lemma}  \label{sec:pro:thm1}
{\rm \bf (Lieb-Thirring estimate)}  
Let $v > 0$ be fixed. There exists a number $L_0$
and a constant $C_M$ (depending on $\Lambda$), 
such that for all
$L \geq  L_0 $ and  $|\triangle| \leq v$ we have
\ben
\left|
\Tr_{L^2(\triangle_c)} ( - \triangle^{\rm M}_{\triangle_c} + V )_- \right|
\leq C_M \int_{\triangle_c} | V_-(x)|^{5/2} \, dx \; ,
\een
where $V$ is any locally integrable
function on $\triangle_c  = \triangle \cap \Lambda_L$ with 
negative part $V_- \in L^{5/2}$.
(Note that $\Tr A_-$ denotes the trace over the negative 
eigenvalues of the selfadjoint operator $A$.)
%
\end{lemma}

\vspace{0.5cm}

\noindent
{\it Proof.} 
We first observe that if $\triangle \cap \Lambda_L = \emptyset$
then 
the estimate is a 
simple consequence of the classical Lieb-Thirring inequality \cite{lie:the}
since we have
Dirichlet boundary conditions on the whole boundary. 
Thus for a given $v$ 
let $L_0$ be as in Lemma  \ref{sec:pro:geom}. Assume that 
$\triangle$ intersects with the boundary of $\Lambda_L$ 
and that 
$|\triangle| \leq v$.  Let 
$\varphi : V \to B_0$  be a  coordinate map with  the properties
as stated in Lemma \ref{sec:pro:geom}. Thus 
$\triangle_c \in V$. We consider first the case where 
$V \cap \Lambda_L = \{ x \in V \ | \ y_3(x) < 0 \}$.
On $B_0$ we define
the reflection 
$\tau : (y_1,y_2,y_3) \mapsto  (y_1,y_2,- y_3)$.
By  $\varphi(\triangle_c)^\tau $ 
we denote the interior of the closure
of   $\varphi(\triangle_c) \cup \tau(\varphi(\triangle_c))$.
Given a function $h$ on $\varphi(\triangle_c)$ we extend it 
to a function $h^{\tau}$ defined a.e. on $\varphi(\triangle_c)^\tau $ by
setting
\ben
h^{\tau}(y) = \left\{ 
\begin{array}{ll}
h(y) \, ,  & \mathrm{if} \ y_3 < 0 \\
 h(\tau(y)) \, ,& \mathrm{if} \ y_3 >  0  \; .
\end{array}
\right.
\een
This establishes the isometric injection
\begin{eqnarray*}
 j : L^2(\varphi(\triangle_c))   &\to& L^2(\varphi(\triangle_c)^\tau) \\
   f     &\mapsto&  2^{-1/2} f^{\tau} \; .
\end{eqnarray*}
Thus for any locally integrable function $W$ we have
\ben
j^* ( - {\Delta}^{\rm D}_{\varphi(\triangle_c)^{\tau}} +   
W^{\tau}) j  
=  - {\Delta}^{\rm M}_{\varphi(\triangle_c)} +  W  \; ,
\een
where $- {\Delta}^{\rm D}_{\varphi(\triangle_c)^{\tau}}$ denotes
the Dirichlet Laplacian on $\varphi(\triangle_c)^{\tau}$ w.r.t.
the Euclidean metric $\delta^{ij}$.
By the Neumann condition, $j$ maps the domain of  
$ - {\Delta}^{\rm M}_{\varphi(\triangle_c)}$ into the domain of 
$- {\Delta}^{\rm D}_{\varphi(\triangle_c)^{\tau}}$.
We then conclude  using Lemma \ref{sec:pro:ineq}
\begin{eqnarray*}
\Tr_{L^2(\triangle_c)}(  - \Delta^{\rm M}_{\triangle_c} +   V )_-  &\geq&
\Tr_{L^2(\triangle_c)}(  - \Delta^{\rm M}_{\triangle_c} +   V_- )_-  \\
& \geq &
C_{\Lambda} \Tr_{L^2(\varphi(\triangle_c))} 
( - {\Delta}^{\rm M}_{\varphi(\triangle_c)} +  C_{\Lambda}^{-1} V_- )_-  \\
&\geq& C_{\Lambda} \Tr_{L^2(\varphi(\triangle_c)^{\tau})}
( - {\Delta}^{\rm D}_{\varphi(\triangle_c)^{\tau}} +   
C_{\Lambda}^{-1} V_-^{\tau})_- \\
&\geq& - C_{\Lambda}^{-3/2} C_{\rm LT} \int_{\varphi(\triangle_c)^{\tau}} 
| V_-^{\tau}(y)|^{5/2} 
\, dy \\
&=& - 2 C_{\Lambda}^{-3/2} C_{\rm LT} \int_{\triangle_c} 
|V_-(x)|^{5/2}  \, dx \; ,
\end{eqnarray*}
where we made abuse of notation by denoting $V_-\circ\phi^{-1}$ by $V_-$.
In the step before last we used the classical Lieb-Thirring 
estimate with constant $C_{\rm LT}$.
If $\Lambda \cap V$ has an edge or a corner the proof
is essentially the same we just  have to perform several reflections,
which affects the value of the 
constant in the inequality by at most
a factor 8, since
in that case, we have to consider the volume obtained by reflecting
$\varphi(\triangle_c)$ on all Neumann planes. Likewise we have to
extend functions defined on $\varphi(\triangle_c)$.
The details are left to the reader.  It follows that the  Lemma holds for 
$C_M = 16 \, C_{\Lambda}^{-3/2} C_{\rm LT}$.
\qed

\vspace{0.3cm}
\noindent
{\it Proof of Lemma \ref{sec:pro:lem2}.}  
Let $\triangle$ be a simplex with $|\triangle| \leq v$. 
Let $L_0$ be sufficiently large such that the assertions
of  Lemmas \ref{sec:pro:geom}
and \ref{sec:pro:thm1} hold. Consider now
$\triangle_c = \triangle \cap \Lambda_L$ for $L \geq L_0$.
The Coulomb interaction is
\begin{eqnarray*}
V_c(x_1,...,x_n,R_1,...,R_k)  & = & \sum_{1 \leq i < j \leq n}
\frac{1}{|x_i - x_j |}  - z \sum_{i=1}^n
\sum_{j=1}^{k} \frac{1}{|x_i - R_j |}  \\
&   & + 
z^2 \sum_{1 \leq i < j  \leq k} \frac{1}{|R_i - R_j |} \; .
\end{eqnarray*}
We introduce the nearest neighbor, or Voronoi, cells $\{ \Gamma_j
\}_{j=1}^k $ defined by
\ben
\Gamma_j = \{ \, x \, | \ |x - R_j | \leq  | x -R_l | \ 
\mathrm{ for \ all} \ l
\neq j \, \} \; .
\een
Furthermore, define the distance $D_j$ of $R_j$ to the boundary of
$\Gamma_j$, i.e.,
\ben
D_j = \mathrm{dist}  ( R_j , \partial \Gamma_j ) = \frac{1}{2} \min
 \{ | R_l - R_j | , \, j \neq l \} \; ,
\een
By Theorem 6 in \cite{lieyau:the}, we have the following inequality 
\begin{equation}  \label{sec:pro:vcx}
V_c ( x_1,..., x_n , R_1 ,... ,R_k ) \geq - \sum_{i=1}^{n} W(x_i) + 
\frac{1}{8} z^2 \sum_{j=1}^k D_j^{-1} \; ,
\end{equation}
where for $x$ in the cell $\Gamma_j$
\ben
W(x) = \frac{2 z + 1}{|x - R_j |} \; .
\een
We note that, in the situation considered here, the coordinates $x_i$
and $R_j$ all lie in 
$\triangle_c$. Using inequality  (\ref{sec:pro:vcx}), we find
\ben
H^{\rm M}_{\bN, \triangle_c} - \mu_n n - \mu_k k \geq 
\sum_{i=1}^n h_i - \mu_k k + \frac{1}{8} z^2 \sum_{j=1}^k D_j^{-1} \; ,
\een
with $h_i = -\Delta^{\rm M}_{\triangle_c, x_i} - W(x_i) - \mu_n$. The
fermion ground state energy of $\sum_{i=1}^n h_i$ is bounded below by
$ 2 \sum_j e_j$, where $e_j$ are the negative eigenvalues of
$h_i$. Hence by   Lemma \ref{sec:pro:thm1} ($f_+(x) = \mathrm{max}
\, ( f(x) , 0 )$)
\begin{eqnarray}
\lefteqn{H^{\rm M}_{\bN, \triangle_c} - \mu_n n - \mu_k k } \nonumber \\
& \geq & - 2 C_M  \int_{\triangle_c} |(W + \mu_n)_+
|^{5/2} dx - \mu_k k + \frac{1}{8} z^2 \sum_{j=1}^{k} D_j^{-1}  \nonumber \\
& \geq & - 2^{5/2}   C_M \int_{\triangle_c} ( |W|^{5/2}
+ | {\mu_n}_+|^{5/2} ) dx - \mu_k k + \frac{1}{8} z^2 \sum_{j=1}^k
D_j^{-1} \; ,  \label{eq:newu}
\end{eqnarray}
where, for the second inequality, we have used that $(a+b)^{5/2} \leq
2^{3/2} ( a^{5/2} + b^{5/2} )$ for $a,b \geq 0$. 
We estimate the first term using  
\begin{eqnarray*}
\int_{\triangle_c} | W|^{5/2} dx & = & \sum_{j=1}^k \int_{\Gamma_j \cap
 \Delta_c} W_j(x)^{5/2} dx  \\
& \leq & \sum_{j=1}^k \int_{|x - R_j | \leq R} (2z + 1)^{5/2} | x -
R_j |^{-5/2} dx  \\
& & + \sum_{j=1}^k 
\int_{ {x \in \Gamma_j \cap \triangle_c} \atop {|x - R_j | \geq R }} 
(2z + 1)^{5/2} R^{-5/2} dx \\
& \leq & (2z + 1)^{5/2} ( 4 \pi k R^{1/2} + | \triangle_c | R^{-5/2} )
\\
& \leq & (2z + 1 )^{5/2} 6 \left( \frac{4 \pi}{5} \right)^{5/6}  |
\triangle|^{1/6} k^{5/6} \; ,
\end{eqnarray*}
where we have made the optimal choice for $R$.
To estimate the term involving the $D_j$, we note that
for $k \geq 2$ ,
\begin{equation}  \label{sec:pro:sd3}
\sum_{j=1}^k D_j^3 \leq \lambda | \triangle | \; ,
\end{equation}
for some constant $\lambda > 0$.
Using H\"olders inequality, i.e.,
\ben
k = \sum_{j=1}^k D_j^{-3/4} D_j^{3/4} \leq \left( \sum_{j=1}^k
  D_j^{-1}     \right)^{3/4}
\left( \sum_{j=1}^k D_j^3    \right)^{1/4} \; ,
\een
we find 
\ben 
k^{4/3} \lambda^{-1/3} |\triangle|^{-1/3} \leq \sum_{j=1}^k 
D_j^{-1}
\een
 for $k \geq 2$. 
Inserting this into (\ref{eq:newu}), we have 
\ben
H^{\rm M}_{\bN, \triangle_c} - \mu_n n - \mu_k k 
\geq - C_1 | \triangle|^{1/6} k^{5/6} 
- C_2 | \triangle | |{\mu_n}_+ |^{5/2} - \mu_k k + 
C_3 |\triangle|^{-1/3} k^{4/3} ( 1 - \delta_{k1})
\een
for some positive constants $0 < C_i < \infty \ , \; 
(i=1,2,3)$, which
depend only on $z$, and $\Lambda$. The case $k=1$ is 
accounted for by $(1 - \delta_{k1} )$. We minimize 
with respect to $k$ with the
result that
\ben
H^{\rm M}_{\bN, \triangle_c} - \mu_n n - \mu_k k \geq 
C_E(\bmu , v  ) \;  ,  \ \forall \ \ |\triangle| \leq v \; ,
\een
for some constant $ C_E(\bmu, v ) \in \R$.
Hence we have shown (a).

\vspace{0.3cm}

To show (b) we decompose the kinetic energy 
\ben
T^{\rm M}_{\triangle_c} = -
\sum_{i=1}^n \Delta^{\rm M}_{ \triangle_c, x_i} - 
(1/M) \sum_{j=1}^k \Delta^{\rm M}_{\triangle_c, R_i}
\een
 and use the same
calculations as in (a). As a  result
\begin{eqnarray*}
\lefteqn{H^{\rm M}_{\bN, \triangle_c} - \mu_n n - \mu_k k } \\ & = &
\frac{1}{2} T^{\rm M}_{\triangle_c} + 
\frac{1}{2} (T^{\rm M}_{\triangle_c}  + 2 V_c + 2 \mu_n n + 2 \mu_k k ) \\
& \geq &  
 \frac{1}{2} T^{\rm M}_{\triangle_c} + \phi(|\triangle|, \bmu , k ) \; ,
\end{eqnarray*}
with
\ben
 \phi(|\triangle|, \bmu , k ) :=
- 2^{3/2} C_1 |
\triangle|^{1/6} k^{5/6} + C_2 | \triangle | |2 {\mu_n}_+|^{5/2} 
- 2 \mu_k k + C_3| \triangle|^{-1/3} k^{4/3} ( 1 - \delta_{k1}) \; .
\een
We estimate the grand canonical partition function as follows
\begin{eqnarray}
\Xi^{\rm M}(\b , \bmu , \triangle_c ) &=&
\Tr_{\HH_{\triangle_c}} e^{- \b ( H^{\rm M}_{\triangle_c} - \bmu \cdot \bN )}
\nonumber \\
&\leq& \sum_{n=0}^{\infty} \Tr_{\bigwedge^n L^2(\triangle_c \times
  \Z_2)} e^{ \b \frac{1}{2} \sum_{i=1}^n \Delta^{\rm M}_{\triangle_c, x_i}}
\nonumber \\
& & \ \times \sum_{k=0}^{\infty} \Tr_{L^2(\triangle_c)^{\otimes k}}
e^{ \b \frac{1}{2M} \sum_{j=1}^k \Delta^{\rm M}_{\triangle_c, R_j}} \cdot
e^{- \b \phi(\Delta, \bmu , k )}  \label{sec:pro:xmb} \; . 
\end{eqnarray} 
If $\triangle$ does not intersect $\Lambda_L$ then we 
have only Dirichlet boundary conditions and in this case 
it is known that the desired bound exists.
It remains to consider the case where  $\triangle$ intersects 
with the boundary of $\Lambda_L$.  
Let
$\varphi : V \to B_0$ be a map with the properties as given in
Lemma \ref{sec:pro:geom}.
We shall first consider the case where  
$V \cap \Lambda_L = \{ x \in V \  |  \ y_3(x) < 0 \ \}$.
We now use the reflection argument and the notation
as introduced in   
the proof of Lemma \ref{sec:pro:thm1}. 
There we have shown that on the form domain of 
$- \triangle^{\rm M}_{\triangle_c}$,
\ben 
- \Delta^{\rm M}_{\triangle_c}  \geq U^* C_{\Lambda} 
( - \Delta^{\rm M}_{\varphi(\triangle_c)} ) U = U^* j^* C_{\Lambda} 
( -  \Delta^{\rm D}_{\varphi(\triangle_c)^{\tau}} ) j U  \; .
\een 
%
Using this estimate we find
\ben
\sum_{n=0}^{\infty}
\Tr_{\bigwedge^n L^2(\triangle_c \times \Z_2)} e^{ \b \frac{1}{2}
  \sum_{i=1}^n \Delta^{\rm M}_{\triangle_c, x_i}} \leq
\sum_{n=0}^{\infty} 
\Tr_{\bigwedge^n L^2(\varphi(\triangle_c)^{\tau} \times \Z_2 )} 
e^{\b \frac{1}{2}
C_{\Lambda} \sum_{i=1}^n \Delta^{\rm D}_{\varphi(\triangle_c)^{\tau}, x_i}} \; .
\een 
The right hand side of this equation is the grand canonical partition
function of an ideal Fermi gas with Dirichlet 
boundary conditions, which is known to be bounded above.
Similarly 
we estimate
\begin{eqnarray*}
 \Tr_{L^2(\triangle_c)^{\otimes k}} e^{ \b \frac{1}{2M} \sum_{j=1}^k
  \Delta^{\rm M}_{\triangle_c, R_j}} & = & 
\left( \Tr_{L^2(\triangle_c)} e^{ \b \frac{1}{2M} 
  \Delta^{\rm M}_{\triangle_c}} \right)^k \\
& \leq & \left( \Tr_{L^2(\varphi(\triangle_c)^{\tau})} e^{ \b \frac{1}{2M} 
  C_{\Lambda} 
\Delta^{\rm D}_{\varphi(\triangle_c)^{\tau}}} \right)^k \\
& \leq & 
\left( \left(\frac{M}{2 \pi \b C_{\Lambda} } \right)^{3/2} 
| \varphi(\triangle_c)^{\tau} | \right)^k
\; ,
\end{eqnarray*}
where the last inequality follows from a standard estimate
\cite{fis:tfe}. 
Note that  
$| \varphi(\triangle_c)^{\tau} | \leq 2 | \triangle|$.
We insert the above inequalities into
eq. (\ref{sec:pro:xmb}). The sum over $k$ converges, 
due to the term with $k^{4/3}$.
Thus we have 
shown (b) for the case where $V \cap \Lambda_L$ does not have
any edges or corners.
If $V \cap \Lam_L$ has an edge or a corner
the proof is essentially the same we just have to perform several 
reflections. We leave the
details to the reader. It turns out
that in the estimates above  $\varphi(\triangle_c)^{\tau}$
is replaced by the volume obtained when reflecting 
$\varphi(\triangle_c)$ on all  Neumann planes. 
Each of the three cases gives us a constant. Taking the 
largest we obtain the desired bound. 
\qed

\subsection{Proof of Theorem \ref{sec:int:thm2}}
\label{sec:pro4}
\label{subsec:34}

As mentioned in Section \ref{sec:modres}, the case $\sigma \geq 0$ is trivial.
Thus let $\sigma < 0$. Everything in the proof of Theorem
\ref{sec:int:thm1} holds if we replace Neumann boundary conditions
with elastic boundary conditions. The only part of the proof 
which  does not generalize trivially 
to elastic boundary conditions is the proof
of Lemma \ref{sec:pro:lem2}. We circumvent this 
by showing that the Laplacian with elastic boundary conditions can
be estimated below in terms of the Laplacian with Neumann boundary
conditions.
We recall that $-\Delta^{\rm M}_{\triangle_c}$ is the unique self adjoint
operator on $L^2(\triangle_c)$ whose quadratic form is the closure
of the form $\phi \mapsto \int_{\triangle_c} | \nabla \phi |^2 \, dx$
with domain $\DD = \{ \, \phi \in H^1(\triangle_c) 
\cap C(\overline{\triangle}_c) \, |
\, \phi \ \mathrm{vanishes \ in \ a \ neighborhood \ of} \  
\partial \triangle
\cap  
\Lambda_L \, \}$. Let $-\Delta^{{\rm M}, \sigma}_{\triangle_c}$  be
the unique self adjoint operator on  $L^2(\triangle_c)$ 
whose quadratic form is the closure
of the form 
\ben
\phi \mapsto \int_{\triangle_c} | \nabla \phi |^2  \, dx
+ \sigma \int_{\triangle \cap \partial \Lambda_L} |\phi|^2 \, dS \; 
\een
with domain $\DD$.
Below we will show that
for all $\triangle_c = \triangle \cap \Lambda_L$, with $|\triangle| \leq v$ 
and $L \geq 1$,
\begin{equation} \label{eq:rest}
- \Delta^{{\rm M},\sigma}_{\triangle_c} \geq \tau 
( - \Delta^{M}_{\triangle_c} ) - C \; ,
\end{equation} 
for some $\tau$, with $0 < \tau \leq 1$, and some finite constant
$C \geq 0$ depending only on $\sigma$ and the geometry of $\Lambda$. 
Thus setting $\boldsymbol{c} = (C,C)$ we have
\begin{eqnarray*}
H^{{\rm M},\sigma}_{\triangle_c} &=& 
 T^{{\rm M},\sigma}_{\triangle_c} + V \\
&\geq&
\tau T^{\rm M}_{\triangle_c} + V  
- \boldsymbol{c} \cdot \bN  \\
&\geq&
\tau^{-1} ( \tau^2 T^{\rm M}_{\triangle_c} + \tau V ) 
- \boldsymbol{c} \cdot \bN  \\
&\cong& 
\tau^{-1} H^{\rm M}_{\tau^{-1} \triangle_c} -    \boldsymbol{c} \cdot \bN  \; .
\end{eqnarray*}
By
\ben
\Tr_{\HH_{\triangle_c}} e^{- \beta(H^{{\rm M},\sigma}_{\triangle_c} - \bmu 
\cdot \bN)}
\leq 
\Tr_{\HH_{\tau^{-1} \triangle_c}} 
e^{- \tau^{-1} \beta(H^{{\rm M},\sigma}_{\tau^{-1} \triangle_c} -   
\tau (\bmu + \boldsymbol{c} ) \cdot \bN ) } \leq C_{\Xi} ( \tau^{-1} \beta,   
\tau ( \bmu + \boldsymbol{c}),
\tau^{-3} v ) \; 
\een
it is now evident that for $\sigma <0$ an  analog of 
Lemma \ref{sec:pro:lem2} 
 for elastic boundary conditions holds.

It remains to show (\ref{eq:rest}). 
Let 
\begin{eqnarray*}
\xi : \overline{\Lambda}  &\to& \R^3  \\
x &\mapsto& \xi(x) = (\xi_1(x),\xi_2(x),\xi_3(x))
\end{eqnarray*}
be a real vector field continuously differentiable in the closed
region $\overline{\Lambda}$ and satisfying the boundary condition
$\boldsymbol{n} \cdot \xi \leq  - 1$ on $\partial \Lambda$ where 
$\boldsymbol{n}$ denotes the inward normal. First observe that
such a vector field exists. 
If $\Lambda$ is a box or has smooth boundary
this is clear.
Consider now the  general case, where the 
boundary of $\Lambda$ has isolated edges and corners.
Since $\Lambda$ is bounded we can cover it with finitely many
sufficiently small open sets $V_{\gamma}$, 
with the property that  
on each of these sets
we can choose
coordinates 
$x \mapsto y(x) = (y_1(x),y_2(x),y_3(x))$ such that  
$\Lambda \cap  V_{\gamma}$
corresponds to either $V_\gamma$, the half space restriction  
$\{x\in V_{\gamma} \ |\ y_3(x)<0 \}$, the quarter 
space restriction $\{x\in V_{\gamma} \ |\ y_2(x),y_3(x)<0 \}$, or the 
octant restriction $\{x\in V_{\gamma} \ |\ y_1(x), y_2(x),y_3(x)<0 \}$,
and such that there exists a vector field on $V_\gamma$
which is  constant in the coordinate
chart and  satisfies the required property on $V_\gamma$. 
Pasting these local vector fields together by means of 
a partition of unity on $\Lambda$ subordinate to the open covering,
we obtain a smooth vector field such that
$\boldsymbol{n} \cdot \xi \leq  - 1$.
Given such a vector field on $\Lambda$, 
then $\xi_L(x) = \xi(x/L)$, for $L \in \R_{+}$,  is a vector field
on $\Lambda_L$ with $\boldsymbol{n} \cdot \xi \leq  - 1$
(here $\boldsymbol{n}$ denotes the inward normal of $\Lambda_L$).
For $\phi \in H^1(\triangle_c) \cap C(\overline{\triangle}_c)$
vanishing in a neighborhood of $\partial \triangle \cap \Lambda_L$
we have 
\begin{eqnarray*}
\int_{\triangle \cap \partial \Lambda_L} | \phi |^2  \, dS 
\leq \int_{\triangle \cap \partial \Lambda_L} 
|\phi|^2 \xi_L ( - \boldsymbol{n} \, dS )
= \int_{\triangle_c} \nabla( \xi_L |\phi|^2) \, dx  \; ,
\end{eqnarray*}
where the equality follows from  Gauss' Theorem.
We calculate
\begin{eqnarray*}
 \nabla( \xi_L |\phi|^2) =  
(\nabla \xi_L )| \phi|^2 +   \xi_L ( \nabla \phi^* )  \phi  +
\xi_L \phi^* ( \nabla \phi) \; ,
\end{eqnarray*}
and for any $\epsilon > 0$ we have
\ben
| \xi_L ( \nabla \phi^* ) \phi  +
\xi_L \phi^* ( \nabla \phi)|  \leq \frac{1}{\epsilon} | \nabla \phi |^2 
+ \epsilon | \xi_L \phi |^2 \; .
\een
As a result
\begin{eqnarray*}
\int_{\triangle \cap \partial \Lambda_L} | \phi |^2  \, dS 
\leq \int_{\triangle_c} \left(
\frac{1}{\epsilon} |\nabla \phi|^2 + \epsilon |\xi_L \phi|^2 +
| \nabla \xi_L | | \phi |^2 \right) \, dx \; .
\end{eqnarray*}
This implies  (\ref{eq:rest}). \qed

\section*{Acknowledgement}

D.H. wants to thank the Department of Mathematics at the University 
of Copenhagen, where this work was started.


\end{document}